\documentclass{article}

\usepackage{PRIMEarxiv}

\usepackage[utf8]{inputenc} % allow utf-8 input
\usepackage[T1]{fontenc}    % use 8-bit T1 fonts
\usepackage{hyperref}       % hyperlinks
\usepackage{url}            % simple URL typesetting
\usepackage{booktabs}       % professional-quality tables
\usepackage{amsfonts}       % blackboard math symbols
\usepackage{nicefrac}       % compact symbols for 1/2, etc.
\usepackage{microtype}      % microtypography
\usepackage{lipsum}
\usepackage{fancyhdr}       % header
\usepackage{graphicx}       % graphics
\graphicspath{{media/}}     % organize your images and other figures under media/ folder

\usepackage{tabularx}       % resize table
\usepackage{multirow}
\usepackage{xcolor}

\title{Uncertainty-Aware Learning Policy for Reliable Pulmonary Nodule Detection on Chest X-ray
}

\author{
  Hyeonjin Choi \\
  Affiliation \\
  Ajou University \\
  Suwon City \\
  \texttt{hjin9122@gmail.com} %%{\{Author1, Author2\}email@email} \\
   \And
  Jinse Kim  \\
  Affiliation \\
  Ajou University \\
  Suwon City \\
  \texttt{jinsae913@gmail.com} \\
   \And
  Dong-yeon Yoo  \\
  Affiliation \\
  Ajou University \\
  Suwon City \\
  \texttt{dongs0125@ajou.ac.kr} \\
  \AND
  Ju-sung Sun \\
  Affiliation \\
  Ajou University Hospital \\
  Suwon City \\
  \texttt{sunnahn@ajou.ac.kr} \\
  \And
  Jung-won Lee \\
  Affiliation \\
  Ajou University \\
  Suwon City \\
  \texttt{jungwony@ajou.ac.kr} \\
}

\begin{document}
\maketitle

\begin{abstract}
    Early detection and rapid intervention of lung cancer are crucial. Nonetheless, ensuring an accurate diagnosis is challenging, as physicians' ability to interpret chest X-rays varies significantly depending on their experience and degree of fatigue. Although medical AI has been rapidly advancing to assist in diagnosis, physicians' trust in such systems remains limited, preventing widespread clinical adoption. This skepticism fundamentally stems from concerns about its diagnostic uncertainty. In clinical diagnosis, physicians utilize extensive background knowledge and clinical experience. In contrast, medical AI primarily relies on repetitive learning of the target lesion to generate diagnoses based solely on that data. In other words, medical AI does not possess sufficient knowledge to render a diagnosis, leading to diagnostic uncertainty. Thus, this study suggests an Uncertainty-Aware Learning Policy that can address the issue of knowledge deficiency by learning the physicians’ background knowledge alongside the Chest X-ray lesion information. We used 2,517 lesion-free images and 656 nodule images, all obtained from Ajou University Hospital. The proposed model attained 92\% (IoU 0.2 / FPPI 2) with a 10\% enhancement in sensitivity compared to the baseline model while also decreasing entropy as a measure of uncertainty by 0.2
\end{abstract}

% keywords can be removed

% Fig.~\ref{fig:fig1}
\section{Introduction}
    Lung cancer is the leading cause of cancer-related deaths, yet it is often diagnosed at stage 3 or later when early intervention is less effective \cite{sung2021global, kratzer2024lung, walters2013lung, ganti2021update}. Improving lung cancer survival depends on accurate diagnosis using chest X-ray (CXR), one of the most widely used imaging modalities worldwide and the initial step in diagnosing chest disease \cite{kamat2015duplication, smith2009radiation, fatihoglu2016x, panunzio2020lung}. Nonetheless, ensuring an accurate diagnosis is challenging, as physicians' ability to interpret CXRs varies significantly depending on their experience and degree of fatigue \cite{turkington2002misinterpretation, del2017missed}. This is why efforts have been made to develop medical AI(Artificial Intelligence) that remains unaffected by these factors. 

    Medical AI is progressing rapidly and achieving superior efficacy in addressing numerous challenges (e.g., diagnostic assistance, pharmaceutical development, prescription management, etc.) \cite{bindra2024artificial, castiglioni2021ai, kaul2020history, younis2024systematic, rajpurkar2022ai}, thereby elevating the expectations of physicians \cite{giavina2024medical, moon2018expectation, doi2005current, chockley2016end}. Nonetheless, a survey found that among physicians and medical students familiar with medical AI, only 10-30\% have used it, suggesting that adoption is still quite low \cite{chen2022acceptance}. Key rationale includes the substantial heterogeneity between datasets due to various inherent factors (e.g., radiologists' expertise, imaging device specifications), the difficulty of ensuring consistent AI model performance across different hospitals, and the lack of clear accountability and countermeasures for misdiagnoses \cite{grzybowski2024challenges, quinn2022three, chen2021artificial, strohm2020implementation}. Due to these worries, a lack of trust in the diagnostic model among healthcare professionals remains a significant obstacle to its adoption. At the core of this skepticism lies the diagnostic uncertainty of AI \cite{salimzadeh2024dealing, seoni2023application, begoli2019need}. In this context, uncertainty refers to how confident a model is in its predictions. When a model exhibits low confidence in its lesion diagnosis, it can be defined with significant diagnostic uncertainty, undermining physicians' trust in its predictions. 

    \begin{figure}
      \centering
      \includegraphics[width=0.9\linewidth]{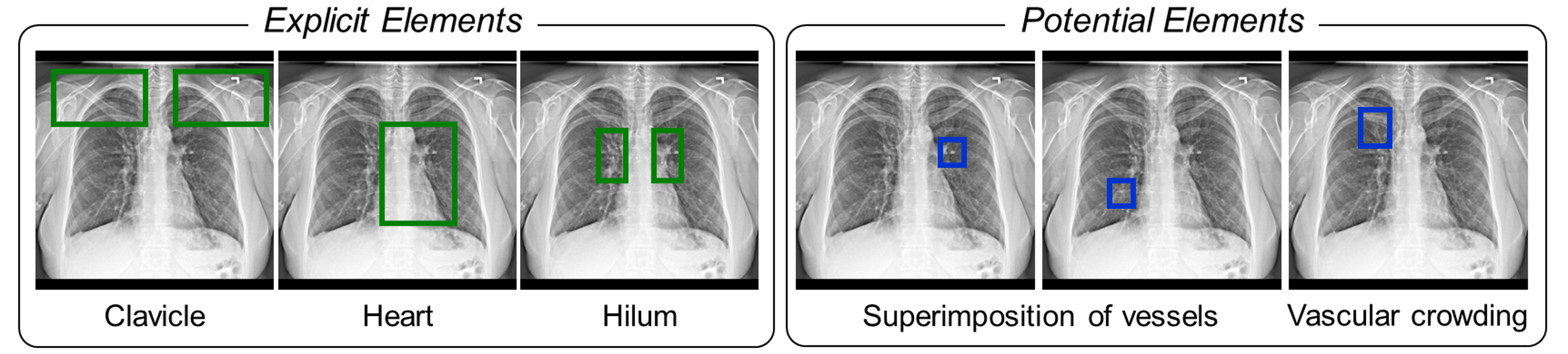}
      \caption{Examples of Physicians’ Extensive Background Knowledge Reflected in CXRs.}
      \label{fig:fig1}
    \end{figure}
    
    Clinicians possess substantial knowledge regarding a CXR before diagnosis in the clinical setting (Fig.~\ref{fig:fig1}). CXR contains explicit elements, such as the lung and heart, and potential elements, including vessel overlap and aggregation, alongside the features of the pulmonary nodule. With this knowledge in mind, physicians apply their experience to enhance the accuracy and confidence of their diagnoses. Conversely, the diagnostic model receives data regarding the target lesion, such as its location and size, and is trained solely on the provided data. These factors can lead the model to wrongly classify entities as nodules, including those that a physician would not deem suitable for such classification. A hilum, for instance, is a basic anatomical structure that has no bearing entirely on a physician’s diagnosis. However, from a model perspective, it can be misinterpreted as an abnormality resembling a nodule in size and brightness. 

    Given this fact, medical images comprise various components readily identifiable to individuals with extensive medical expertise yet pose challenges for those lacking such knowledge, resulting in diagnostic uncertainty due to possible misinterpretation. Enhancing physicians' trust in medical AI requires overcoming this knowledge gap by including their background knowledge in the learning model alongside lesion information, thus resolving the cause of contributing to diagnostic uncertainty. Therefore, this study proposes an uncertainty-aware learning policy that addresses the knowledge deficiency in CXR by supplying background knowledge (explicit \& potential factors) and lesion information.  

    The proposed policy was evaluated using 2,517 lesion-free images and 656 nodule images obtained from Hospital A. In line with this policy, we built and evaluated the most popular detection model, YOLOv7 \cite{wang2023yolov7}, which achieved a recall of 92\% (IoU 0.2 / FPPI 2), a 10\% improvement over the prior model. This improvement suggests that the model reduces ambiguity regarding other variables resembling nodules due to insufficient knowledge, enhancing confidence in nodule diagnosis. The proposed policy is quantitatively verified to reduce the entropy of diagnostic results by 0.2, thereby confirming its effectiveness as a strategy for improving uncertainty. In summary, the contributions of the suggested strategy are outlined as follows: 

    \begin{itemize}
        \item[1.] The proposed learning policy incorporates physicians' diagnostic expertise into the model and offers a pragmatic approach that may enhance trust in AI-assisted diagnoses. 
        \item[2.] The proposed learning policy enhances the model with supplementary knowledge, mitigating misdiagnosis from elements that may be mistaken for the target lesion, thus improving diagnostic performance. 
        \item[3.] The proposed learning policy is independent of the model, allowing it to apply to any detection model architecture. 
        \item[4.] A pulmonary nodule diagnostic model developed with the proposed policy functions as a highly reliable and high-performance tool, rendering it appropriate for use as a secondary opinion in clinical practice. 
    \end{itemize}

\section{Methods}
\label{method}

    \begin{figure}
      \centering
      \includegraphics[width=0.95\linewidth]{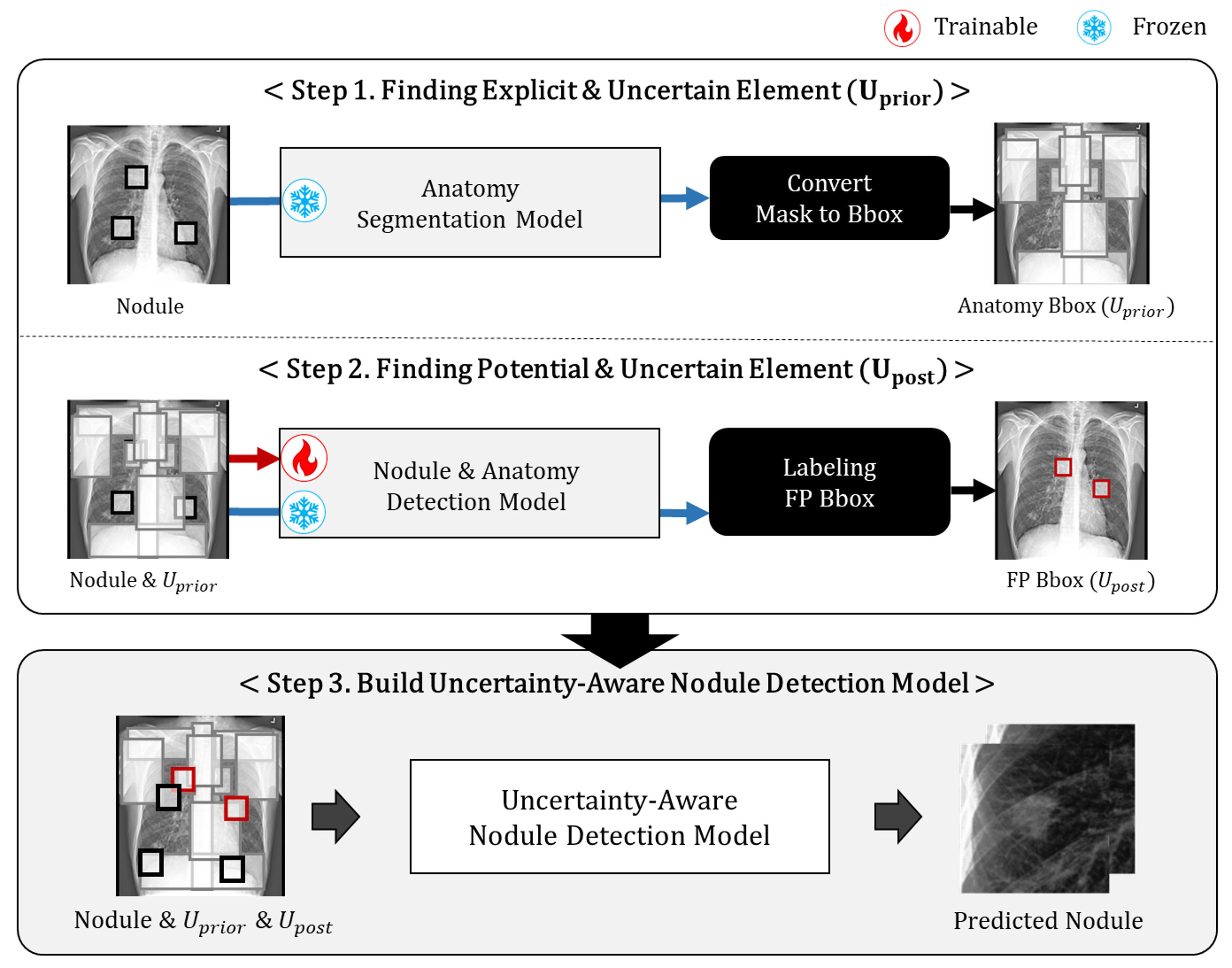}
      \caption{\textbf{Uncertainty-Aware Learning Process.} This figure illustrates the training process, where the model sequentially receives $U_{prior}$ and $U_{post}$ extracted and labeled from the nodule dataset. In the figure, 'trainable' indicates that the model is in training mode, meaning its parameters are updated. Meanwhile, 'frozen' denotes inference mode, where the model is used for prediction without parameter updates.}
      \label{fig:fig2}
    \end{figure}

    \subsection{Analysis of Uncertainty Sources}
        We separate the two categories of physician background knowledge, explicit and potential factors, into predefined and detectable factors ($U_{prior}$) and undefined and undetectable factors ($U_{post}$). In this context, "predefined and detectable" implies that each label can be established independently of the nodule detection model. First, the explicit component comprises anatomical structures (e.g., heart, clavicle, etc.), essential for comprehending CXRs despite lacking similarities to nodules. Their presence remains consistent in location and shape across individuals, enabling developers with relevant expertise to perform labeling independently. Open segmentation models are also capable of extracting relevant features. Thus, the explicit element is considered an uncertainty element, $U_{prior}$, which is detectable and predefined independent of the nodule detection model. Potential elements refer to inadvertently captured features, such as overlapping vessels or vessel clustering. These features are not uniformly observable; however, they exhibit a nodule-like appearance on the CXR, and this information should be incorporated into the model to mitigate diagnostic ambiguity. Potential elements are incidental characteristics, such as overlapping vessels or vessel clustering, that are not uniformly observable. On the CXR, however, they appear like nodules, so the details should be fed into the model to clear up any diagnostic ambiguity. Most false positives(FP) in nodule detection models stem from potential elements like vessel overlap. However, these elements do not have a consistent location across individuals, and their size and brightness vary, making labeling difficult even with relevant expertise. Moreover, no open-access model has been introduced. A nodule detection model must first be developed to identify which elements in the training data are misclassified as nodules. Thus, we define potential elements as predefined and indiscernible elements $U_{post}$, classified based on the nodule detection model. 
    
    \subsection{Uncertainty-Aware Learning Policy}
        The proposed learning policy is illustrated in Fig.~\ref{fig:fig2}. It demonstrates constructing and supplying a labeled dataset with lesion information, enabling it to assimilate the knowledge underlying a physician's nodule diagnosis process. Employ an open segmentation model to acquire a mask for each component of the anatomical structure and extract the Bounding Box (Bbox) data. This constitutes the $U_{prior}$ data. Subsequently, construct a detection model utilizing the nodule data alongside the $U_{prior}$ data. Inference is subsequently conducted on the training data to acquire the FP Bbox data, precisely the elements in the CXR that the nodule detection model misidentifies as nodules. The FP Bbox instances are annotated and classified as $U_{post}$ data. The completed $U_{prior}$ and $U_{post}$, along with the nodule data, are input into the nodule detection model to develop a reliable detector that incorporates diagnostic uncertainty, thereby enhancing the model's knowledge gap from a physician's viewpoint.

\section{Experiment and Results}
    \subsection{Dataset Preparation}
        \textbf{Dataset.} We employed one private dataset and three open-access datasets. The private dataset was supplied by Ajou University Hospital, Suwon, Republic of Korea. And comprises 659 nodule CXRs and corresponding nodule mask data. We used an 8:2 split for training and validation while developing the suggested learning policy-based nodule detection model. The three open datasets, CHN, MCU, and Indiana \cite{stirenko2018chest, jaeger2014two, xue2023cross}, comprise lung region masks and contain 566, 138, and 55 pairs of CXRs and lung masks, respectively. They were used to verify the anatomy segmentation model's functionality. 

        \textbf{Preparing Nodule and $U_{prior}$ Data.}  To validate our proposed learning policy, we construct a dataset of bounding boxes of nodules and predefined data $U_{prior}$. The bounding boxes of nodules are derived from the maxima and minima of the white pixel coordinates along the x and y axes in our nodule mask data. The bounding boxes of the individual components of $U_{prior}$ were derived from the mask data using an open segmentation model and extracted similarly to the nodules. 

    \subsection{Experimental Setup}
        \textbf{Evaluation Metric.} We employ the Intersection over Union (IoU), Precision-Recall (PR) Curve, Free-Response Receiver Operating Characteristic (FROC) Curve, and Area Under Curve (AUC) of both curves as assessment metrics. The IoU serves as a key metric for assessing detection model performance, which we employed to determine the proposed learning policy's prediction accuracy intuitively. PR Curve and FROC are metrics to evaluate the variation in sensitivity concerning FPs. While assessing the PR Curve and its AUC is common, the precision metric is not readily comprehensible to users in terms of the number of FPs it signifies. Consequently, we employed the FROC metric, which indicates sensitivity relative to the average number of FPs Per individual Image (FPPI). 

        \textbf{Anatomy Detection Model Validation} We assessed the efficacy of an anatomical structure segmentation model from the open library TorchXrayVision \cite{cohen2022torchxrayvision}, which we employed to construct the $U_{prior}$ dataset. The model generates masks for ten anatomical structures: clavicle, scapula, lung, hilum, heart, aorta, diaphragm, mediastinum, trachea, and spine, derived from CXR. Comprehensive validation of performance requires segmentation data for individual elements. However, mask data for anatomical structures beyond the lungs is not publicly accessible. Further, this study aims to use the model to create a detection dataset. Consequently, the model's performance was assessed based on its detection prediction efficacy across three publicly available lung mask datasets. The findings demonstrate reliable performance, achieving an average IoU exceeding 0.93. 

        \textbf{Implementation Details.} The detection model employed to validate the proposed learning policy utilized the default YOLOv7. The learning parameters were trained for 200 epochs using the default values without modification. In all experiments, the input image dimensions were standardized to 512x512.
    
    \subsection{Uncertainty-Aware Learning Policy Validation}
        \textbf{Batch Size Selection for Efficient Training Nodule Detection} Empirically ascertain the optimal batch size for nodule training. Upon training the nodule data for batches 2, 4, and 8, the PR Curve AUC for each model at IoU 0.5 was 0.698, 0.661, and 0.553, respectively, with batch 2 exhibiting the highest AUC. We established the batch size at 2 for the remaining experiments. 
    
        \begin{figure}
          \centering
          \includegraphics[width=\textwidth]{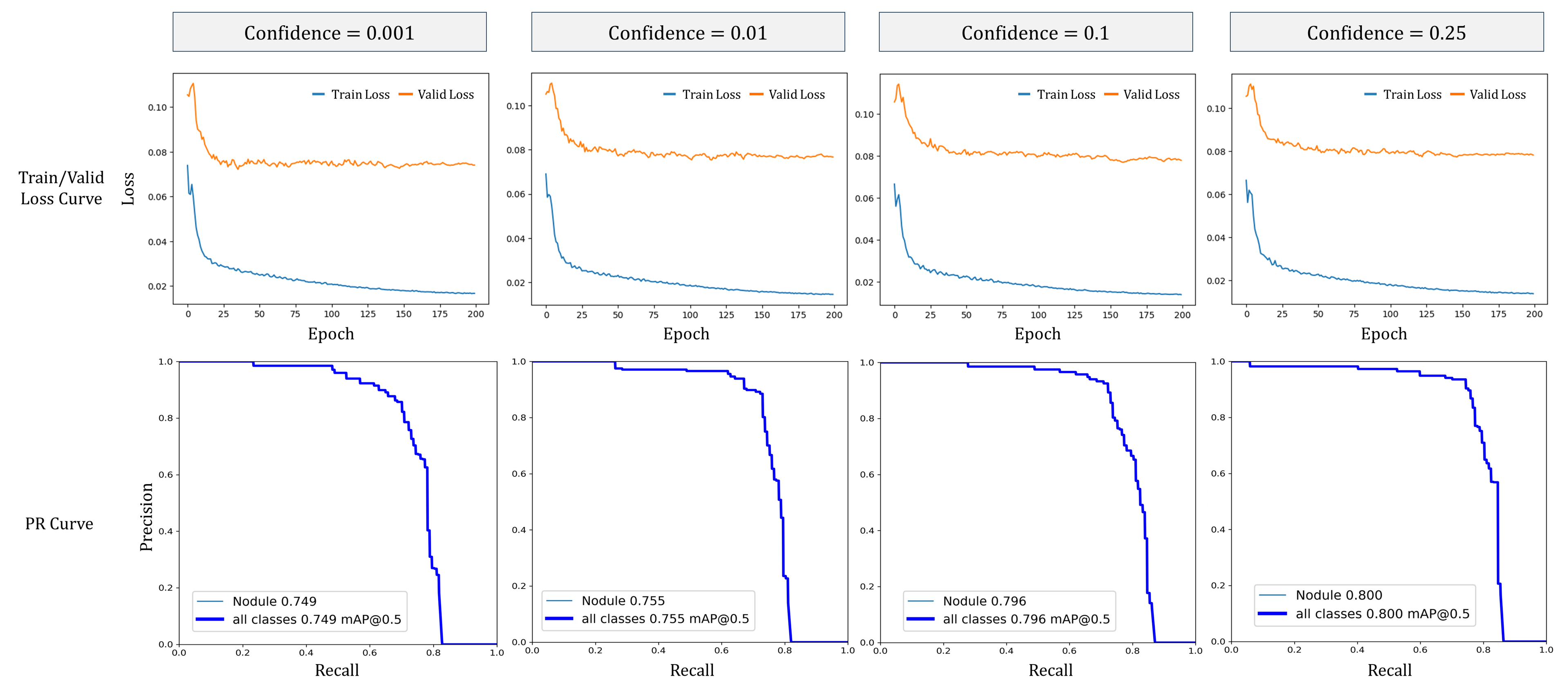}
          \caption{The Results of FP-based Learning Under Different Confidence Values.}
          \label{fig:fig3}
        \end{figure}

        \textbf{Confidence Threshold Calibration for Optimal $U_{post}$ Data Construction.} Establish the FP labeling threshold experimentally for the construction of the $U_{post}$ dataset. FP data labeling involves isolating the incorrect prediction (FPs) solely from the data estimated by the model as nodules. The model's prediction is contingent upon the confidence threshold. Setting a high confidence threshold will result in only those predictions in which the model exhibits strong confidence being designated as nodules. In this instance, learning may prove challenging due to the lack of $U_{post}$ data. Conversely, if the threshold is set excessively low, slightly ambiguous predictions may be classified as nodules, leading to a substantial increase in FP. The model may prioritize false positives over nodules. When the confidence was adjusted to YOLOv7's training default of 0.25, there were 46 FPs. However, when set to the validation default of 0.001, the count escalated to 3,042 FPs, indicating a significant number disparity with nodule data. Fig.~\ref{fig:fig3} illustrates the learning outcomes of the FP-based nodule detection model developed under differing confidence levels. The learning model utilizing FPs identified at 0.25 exhibits optimal performance, suggesting that the $U_{post}$ was constructed with an appropriate volume of data. Therefore, we established confidence at 0.25. 

        \begin{figure}
          \centering
          \includegraphics[width=0.9\linewidth]{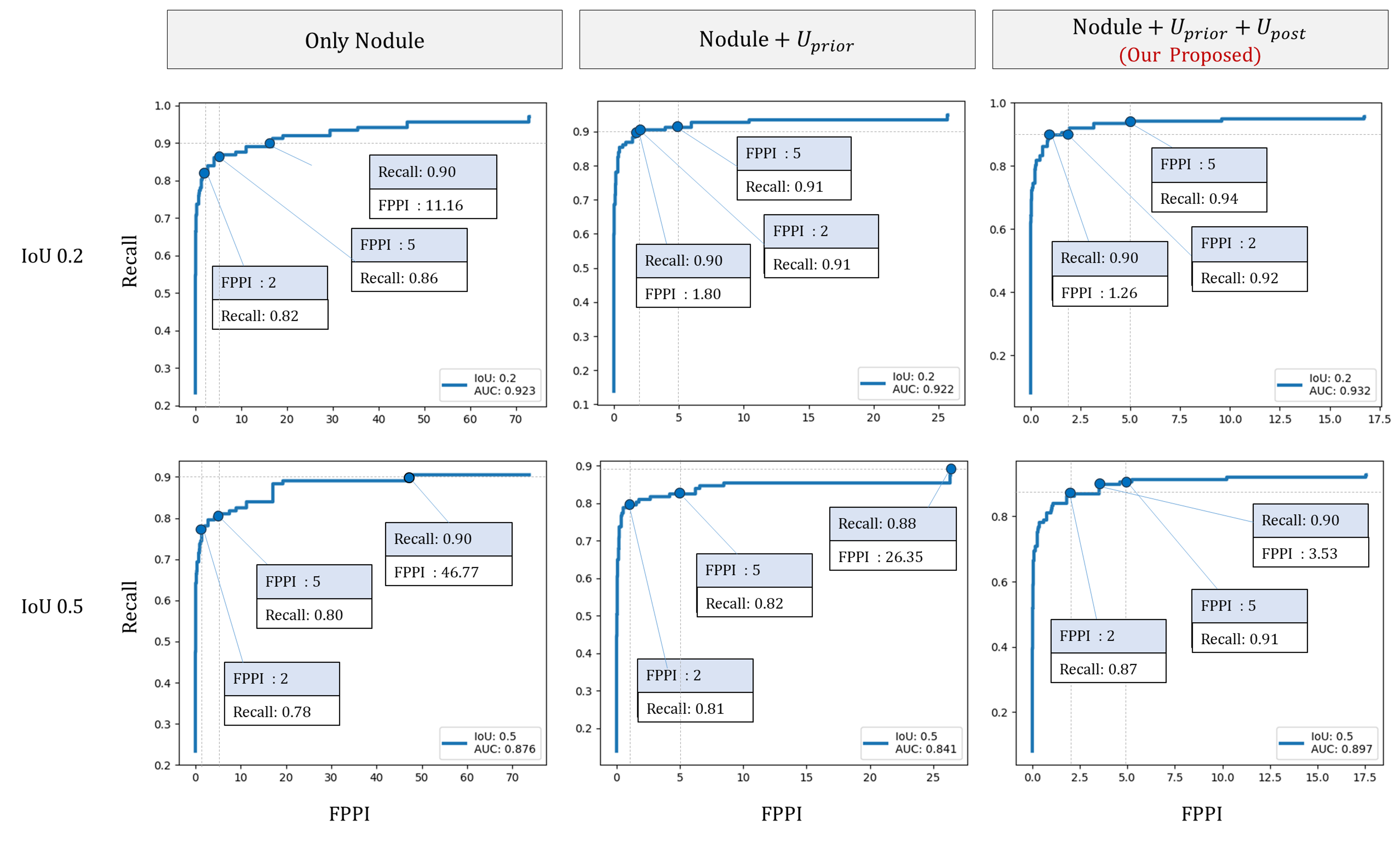}
          \caption{The Results of FP-based Learning Under Different Confidence Values.}
          \label{fig:fig4}
        \end{figure}

        \textbf{Learning Policy Validation through Ablation Study.} Ultimately, we assess the diagnostic reliability of the proposed learning policy through an ablation study involving three models: a nodule model, a model trained on nodules, and $U_{prior}$, and a model trained on nodules, $U_{prior}$, and $U_{post}$. Fig.~\ref{fig:fig4} illustrates the FROC curves of the three models at IoU values of 0.2 and 0.5. The IoU values of 0.2 and 0.5 are standard metrics. The figure demonstrates that, in both scenarios, our proposed model exhibits superior performance. The sensitivities of the three models are 0.82, 0.91, and 0.92, respectively, corresponding to an IoU of 0.2 and an FPPI of 2. The proposed model exhibits optimal performance with an average of 2 FPPI. The results indicate that the anatomical structure model and the proposed model exhibit comparable performance. Nevertheless, the proposed model demonstrates a sensitivity of approximately 0.3 higher at an FPPI of 5. Furthermore, it converges to values exceeding 0.9 for lower FPPIs, indicating that greater sensitivity is attained with increased reliability. At an IoU of 0.5, the proposed model demonstrates markedly superior performance at the same FPPI level compared to an IoU of 0.2. This suggests that both the nodule model and the anatomical structure model exhibit limitations in accurately pinpointing the nodule's precise location. The results indicate that educational approaches supplying physicians' extensive background knowledge can significantly enhance the performance of nodule detection models. 

        \begin{figure}
          \centering
          \includegraphics[width=0.8\linewidth]{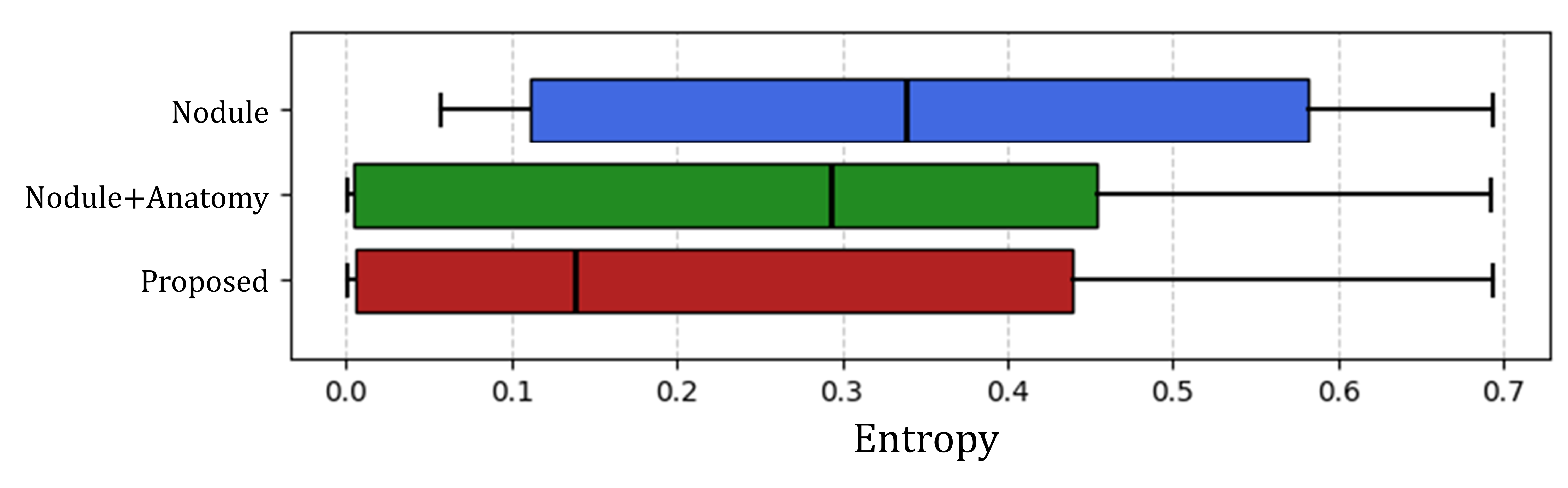}
          \caption{The Results of FP-based Learning Under Different Confidence Values.}
          \label{fig:fig5}
        \end{figure}

        \begin{equation}
            Entropy(p) = -(p\log p+(1-p)\log (1-p))
        \end{equation}
        
        \textbf{Evaluation of Uncertainty Reduction Based on Entropy.} We quantitatively assessed the enhancement in predictive confidence of the proposed learning policy utilizing the traditional uncertainty quantification metric, Entropy. Equation (1) represents the formula for entropy. In the formula, p represents the confidence of a singular prediction, and entropy decreases as the confidence of that prediction increases. Fig.~\ref{fig:fig5} illustrates a boxplot depicting the entropy distribution of individual predictions on the validation dataset for the three models previously assessed in the ablation study. The entropy of the models employing our proposed learning policy is markedly lower than that of the baseline model, which is learned solely from nodules. The median entropy values are approximately 0.3388, 0.2935, and 0.1383, suggesting that the proposed method decreases entropy by approximately 0.2.

\section{Conclusion}
    This study proposes a learning policy to enhance pulmonary nodule detection models' reliability by including physicians' expertise in the diagnostic process. A learning policy was developed by categorizing the physician's knowledge into explicit and potential components and employing a labeling method specific to each attribute, enabling it to learn alongside pulmonary nodule data. 
    
    The detection model developed with the proposed learning policy attained 92\% sensitivity (IoU 0.2 / FPPI 2), reflecting a 10\% enhancement over the baseline model and reducing entropy as a measure of uncertainty by 0.2. We have demonstrated that our proposed learning policy enhances model uncertainty while improving performance. Also, supplying the model with diagnostic background knowledge emulates the clinical diagnostic process, thereby earning the trust of clinicians. urthermore, given that the proposed policy is entirely independent of the model architecture, it is anticipated to be scalable across diverse detection challenges. The pulmonary nodule diagnostic model developed with the proposed policy is a dependable and high-performance tool suitable for use as a second opinion in clinical practice. 
    
    This study was constrained by the limited availability of pulmonary nodule Bbox data, resulting in validation experiments being conducted solely on a single private dataset. Nonetheless, as previously indicated, it is anticipated that optimal performance can be attained without limitations on the data domain or model architecture. Therefore, we will undertake validation experiments to assess the proposed learning policy's scalability by acquiring future supplementary data. 

\section*{Acknowledgments}
This work was supported by the IITP(Institute of Information \& Coummunications Technology Planning \& Evaluation)-ITRC(Information Technology Research Center) grant funded by the Korea government(Ministry of Science and ICT)(IITP-2025-RS-2020-II201461)

%Bibliography
\bibliographystyle{unsrt}  
\bibliography{references}

\end{document}